\documentclass[11pt,a4paper]{article}

\usepackage{graphicx}
\usepackage{color}
\usepackage{amsmath,amssymb}
\usepackage{bbold}
\usepackage{t1enc}
\usepackage{cite}

\DeclareMathSymbol{\shortminus}{\mathbin}{AMSa}{"39}
\DeclareMathSymbol{\shm}{\mathbin}{AMSa}{"39}
\newcommand{\oh}{\textstyle \frac{1}{2}}
\newcommand{\moh}{\textstyle \shortminus \!\! \frac{1}{2}}

\begin{document}

\begin{center}
\begin{Large}
{\bf Post-decay quantum entanglement in top pair production}
\end{Large}

\vspace{0.5cm}
\renewcommand*{\thefootnote}{\fnsymbol{footnote}}
\setcounter{footnote}{1}
J.~A.~Aguilar--Saavedra \\[1mm]
\begin{small}
Instituto de F\'isica Te\'orica IFT-UAM/CSIC, c/Nicol\'as Cabrera 13--15, 28049 Madrid, Spain \\
\end{small}
\end{center}

\begin{abstract}
Top pairs produced at the Large Hadron Collider exhibit quantum entanglement of their spins near threshold and for boosted, central $t \bar t$ pairs. The entanglement is maintained between the decay products, in particular between the top quark and the $W^-$ boson from the anti-quark (or vice-versa, between $\bar t$ and $W^+$) in certain kinematical regions. Therefore, $t \bar t$ production provides a  rare opportunity to verify the spin entanglement between a fermion and a boson. The $tW$ entanglement can be probed at the $7\sigma$ level near threshold with Run 2 data, and at the $5\sigma$ level in the boosted region with the foreseen Run 3 luminosity. In addition, the entanglement between the two $W$ bosons can be probed at the $4\sigma$ level at the LHC Run 3.

\end{abstract}

\section{Introduction}

Quantum mechanics is one of the fundamental pillars of modern particle physics and, as such, testing it thoroughly is of the utmost importance. 
Quantum entanglement can be tested at the energy frontier in $pp$ collisions at the Large Hadron Collider (LHC). Proposals have been made for $t \bar t$ production~\cite{Afik:2020onf,Fabbrichesi:2021npl,Severi:2021cnj,Afik:2022kwm,Aguilar-Saavedra:2022uye,Afik:2022dgh,Dong:2023xiw}, and a preliminary measurement has been performed by the ATLAS Collaboration~\cite{note}. Entanglement can be also tested for vector bosons from Higgs decays~\cite{Barr:2021zcp,Aguilar-Saavedra:2022wam,Aguilar-Saavedra:2022mpg} and electroweak diboson production~\cite{Ashby-Pickering:2022umy,Fabbrichesi:2023cev,Morales:2023gow}. In all these cases, and other proposals for future colliders~\cite{Altakach:2022ywa} the entanglement takes place between the spins of the produced particles, which are either fermion pairs ($t \bar t$, $\tau^+ \tau^-$) or boson pairs ($WW$, $ZZ$, $WZ$). Top pair production also offers the rare and exciting possibility to test the spin entanglement between a fermion and a boson, namely the top quark and the $W^-$ from the $\bar t$ decay (or their charge conjugate). Previous tests of fermion-boson entanglement have been provided in Ref.~\cite{Feist:2022zwe}, where electron-photon entangled pairs have been achieved by creating photons off an electron beam. Moreover, it allows to test the {\em post-decay} entanglement: the coherence of the top pair is propagated to its decay products and can be observed in certain kinematical regions. And so, it manifests that the decay of a particle is not a {\em measurement} in the quantum-mechanical sense.

In order to understand how the $tW$ entanglement arises let us for example consider $t \bar t$ production from gluon fusion near threshold. The spin state is approximately a singlet
\begin{equation}
\psi_{t \bar t} = \frac{1}{\sqrt 2} \left[ |\oh \, \moh \rangle - |\moh \, \oh \rangle \right]  \,,
\end{equation}
where we take the $\hat z$ spin quantisation axis in the direction of one proton, $\hat z = \hat p_p \equiv (0,0,1)$ for definiteness. In general, the coherence between the $t$ and $W^-$ spins is lost upon integration over the $\bar t$ decay phase space and sum over $\bar b$ polarisations which are difficult, if not impossible, to measure. However, let us assume that the $W^-$ three-momentum direction $\vec p_W$ in the $\bar t$ rest frame is close to the $\hat z$ axis (and therefore the $\bar b$ three-momentum approximately in the $-\hat z$ direction). The left-handed $tbW$ interaction mediating the top quark decay produces a left-chirality $\bar b$, therefore, up to small $m_b/m_t$ effects, the $\bar b$ quark has positive helicity, i.e. it is in a $S_z = - \oh$ state. For a $S_z = - \oh$ top anti-quark this implies $S_z = 0$ for the $W^-$ boson, because orbital angular momentum in the direction of motion vanishes. Conversely, for a $S_z = \oh$ top anti-quark it implies $S_z = +1$ for the $W^-$. Thus, the spin state of the $tW$ pair is
\begin{equation}
\psi_{t W^-} = \frac{1}{\sqrt{a^2 + b^2}} \left[ a |\oh \, 0 \rangle - b |\moh \, 1 \rangle \right]  \,,
\label{ec:tWvec}
\end{equation}
where $a$ and $b$ are not expected to be equal because of polarisation effects in the $\bar t$ decay: the $W^-$ angular distribution in the $\bar t$ rest frame is not isotropic. If, instead, we consider $\vec p_W$ in the $-\hat z$ direction, the spin state is
\begin{equation}
\psi'_{t W^-} = \frac{1}{\sqrt{{a}^2 + {b}^2}} \left[ b |\oh \, -1 \rangle - a |\moh \, 0 \rangle \right]  \,.
\end{equation}
In practice, it is sufficient to consider some relatively wide interval for the angle $\theta_W$ between $\vec p_W$ and either the positive or negative $\hat z$ axis to experimentally verify the entanglement.

\section{Theoretical setup}
\label{sec:2}

For a system composed of two subsystems $A$ and $B$, a mixed state is said to be separable if the density operator describing this state can be written in the form
\begin{equation}
\rho = \sum_n \, p_n \rho_n^A \otimes \rho_n^B \,,
\label{ec:rhosep}
\end{equation}
where $\rho_n^{A,B}$ are density operators for the two subsystems $A$, $B$, respectively, and $p_n$ are classical probabilities, $p_n \geq 0$ with $\sum_n p_n = 1$. If $\rho$ cannot be written in this fashion, the state is said to be entangled. A necessary condition for the state to be separable is given by the Peres-Horodecki criterion~\cite{Peres:1996dw,Horodecki:1997vt}: taking the transpose of the density operator in one of the two subspaces, e.g. in the $B$ space, the resulting density operator $\rho^{T_2}$ must still be valid, in particular with non-negative eigenvalues. This condition can be understood since, if $\rho$ is expressed as in (\ref{ec:rhosep}), the transpose of $(\rho_n^B)^T$ is still a valid density operator for $B$, therefore $\rho^{T_2}$ is a valid density operator with non-negative eigenvalues.
The Peres-Horodecki criterion is also a sufficient condition when the dimensions of the Hilbert spaces are $\text{dim}~\mathcal{H}_A = \text{dim}~\mathcal{H}_B = 2$, or $\text{dim}~\mathcal{H}_A = 2$, $\text{dim}~\mathcal{H}_B = 3$. We therefore use as `entanglement indicator' the lowest eigenvalue of $\rho^{T_2}$,
\begin{equation}
\lambda_1 \equiv \operatorname{min} \{ \lambda_i \} \,.
\end{equation} 
When $\lambda_1 < 0$, this is a sufficient condition for entanglement. However, we point out that even when all $\lambda_i$ are positive, statistical fluctuations may result in negative eigenvalues when measuring $\rho^{T_2}$ in data. The bias associated to this effect is discussed, and corrected for, in section~\ref{sec:4}.

We parameterise the top quark and $W$ boson density matrices using irreducible tensor operators. For the top quark we use
\begin{align}
& t_1^1 = - \frac{1}{\sqrt 2} (\sigma_1 + i \sigma_2) \,, \quad t_{-1}^1 = \frac{1}{\sqrt 2} (\sigma_1 - i \sigma_2) \,,\quad t_0^1 = \sigma_3 \,,
\end{align}
with $\sigma_i$ the Pauli matrices. For the $W$ boson we use an analogous definition for the $L=1$ operators but with different normalisation,
\begin{align}
& T_1^1 = - \frac{\sqrt 3}{2} (S_1 + i S_2) \,, \quad T_{-1}^1 = \frac{\sqrt 3}{2} (S_1 - i S_2) \,, \quad T_0^1 = \sqrt{\frac{3}{2}} S_3 \,,
\end{align}
with $S_i$ the spin-1 operators in the Cartesian basis. The $L=2$ operators are~\cite{Aguilar-Saavedra:2015yza}
\begin{align}
& T_{\pm 2}^2 = \frac{2}{\sqrt 3} (T_{\pm 1}^1)^2 \,, \quad 
T_{\pm 1}^2 = \sqrt{\frac{2}{3}} \left[ T_{\pm 1}^1 T_0^1 + T_0^1 T_{\pm 1}^1 \right] \,, \quad \notag \\
&  T_0^2 = \frac{\sqrt 2}{3} \left[ T_1^1 T_{-1}^1 + T_{-1}^1 T_1^1 + 2 (T_0^1)^2 \right] \,.
\end{align}
The operators satisfy $(t^1_m)^\dagger = (-1)^m t^1_{-m}$, $(T^L_M)^\dagger = (-1)^M T^L_{-M}$ and their normalisations are chosen so that $\operatorname{tr} [ t_{m_1}^1 (t_{m_2}^1)^\dagger ] = 2 \delta_{m_1 m_2}$, $\operatorname{tr} [ T_{M_1}^{L_1} (T_{M_2}^{L_2})^\dagger ] = 3 \delta_{L_1 L_2} \delta_{M_1 M_2}$. The density operator of the $tW$ pair can then be parameterised as
\begin{align}
\rho_{tW} = \frac{1}{6} \left[ \mathbb{1}_2 \otimes \mathbb{1}_3 + a_m \, t^1_m \otimes \mathbb{1}_3 + A_{LM} \, \mathbb{1}_2 \otimes T^L_M + C_{mLM} \, t^1_m \otimes T^L_M  \right] \,,
\label{ec:ptW}
\end{align}
where a sum over repeated indices $m$, $L$, $M$ is understood. The constants $a_m$ and $A_{LM}$ are the top and $W$ boson polarisations, respectively, and $C_{mLM}$ are their spin correlations. These coefficients satisfy
\begin{equation}
a_{-m} = (-1)^m a_m^* \,, \quad A_{L \,\shm M} = (-1)^M A_{LM}^* \,, \quad
C_{\shm m L \,\shm M} = (-1)^{m+M} C_{mLM}^* \,.
\end{equation}
Therefore, $a_0$, $A_{L0}$ and $C_{0L0}$ are real and the remaining coefficients are in general complex. 
We note that many previous studies for top polarisation use a parameterisation in terms of Pauli matrices and a real polarisation vector $P^\text{\tiny CAR}$ in Cartesian coordinates, with
\begin{equation}
P^\text{\tiny CAR}_1 = \frac{1}{\sqrt 2} (-a_1 + a_{\shm 1}) \,,\quad P^\text{\tiny CAR}_2 = -\frac{i}{\sqrt 2} (a_1 + a_{\shm 1}) \,, \quad P^\text{\tiny CAR}_3 = a_3 \,.
\end{equation}
We prefer to use a polar basis with a complex polarisation vector $a_m$, so that the treatment of the top quark and $W$ boson is more alike. The explicit expressions of $\rho_{tW}$ and $\rho_{tW}^{T_2}$ are given in appendix~\ref{sec:a}.

The different terms in the density operators are not directly accessible, but can be measured via the angular distributions of the top and $W$ decay products, which are used as {\em spin analysers}. 
In our case, it is best to use the charged leptons $\ell = e,\mu$ from the $t \to W^+ b \to \ell^+ \nu b$ and $W^- \to \ell^- \nu$ decays, so for simplicity we particularise the otherwise general framework to this specific case. Let us label the three-momentum direction of $\ell^+$ in the top quark rest frame as
$\hat p_1$, and the three-momentum direction of $\ell^-$ in the $W^-$ rest frame as $\hat p_2^*$. (The asterisk highlights the fact that the $\ell^-$ three-momentum is taken in the $W$ rest frame.) In polar coordinates,
\begin{align}
\hat p_1 = (\sin \theta_1 \cos \varphi_1, \sin \theta_1 \sin \varphi_1, \cos \theta_1 ) \,, \notag \\
\hat p_2^* = (\sin \theta_2^* \cos \varphi_2^*, \sin \theta_2^* \sin \varphi_2^*, \cos \theta_2^* ) \,.
\end{align}
The $(\hat x,\hat y,\hat z)$ reference system, whose orientation is necessary to define the angles $\Omega_1 = (\theta_1,\varphi_1)$ and $\Omega_2^* = (\theta_2^*,\varphi_2^*)$, is the same one used to write the density operator. The charged lepton momenta in the $t$ and $W^-$ frames have to be obtained with a succession of boosts, see for example Ref.~\cite{Aguilar-Saavedra:2022kgy} for a detailed discussion.

The decay distributions can be obtained by convoluting the density operator with the appropriate decay angular density matrices~\cite{Rahaman:2021fcz}. For the top quark, the decay density matrix is
\begin{equation}
\Gamma_1 = \frac{1}{2} \left( \! \begin{array}{cc}
1 + \alpha \cos \theta_1 & \alpha \sin \theta_1 e^{i \varphi_1} \\
\alpha \sin \theta_1 e^{-i \varphi_1} & 1 - \alpha \cos \theta_1
\end{array} \! \right) \,,
\end{equation}
with $\alpha = 1$ for the positive charged lepton. The $W-$ decay density matrix is
\begin{footnotesize}
\begin{equation}
\Gamma_2 =\frac{1}{4}
\left(\!\!
\begin{array}{ccc}
 1 + \cos^2 \theta_2^* -2 \eta_\ell \cos \theta_2^* 
 & \frac{1}{\sqrt{2}} (\sin 2 \theta_2^* - 2 \eta_\ell \sin\theta_2^* ) e^{i \varphi_2^*}
 & (1 - \cos^2 \theta_2^*) e^{i 2 \varphi_2^*} \\
 \frac{1}{\sqrt{2}} (\sin 2 \theta_2^* - 2 \eta_\ell \sin \theta_2^*) e^{-i \varphi_2^*}
 & 2 \sin^2 \theta_2^*
 & -\frac{1}{\sqrt{2}} (\sin 2 \theta_2^* + 2 \eta_\ell \sin \theta_2^*) e^{i \varphi_2^*}  \\
 (1 - \cos^2 \theta_2^*) e^{-i 2 \varphi_2^*}
 & -\frac{1}{\sqrt{2}} (\sin 2 \theta_2^* + 2 \eta_\ell \sin \theta_2^*) e^{-i \varphi_2^*}
 & 1+\cos^2 \theta_2^* - 2 \eta_\ell \cos \theta_2^* \\
\end{array} \!\! \right) \,.
\end{equation}
\end{footnotesize}
with $\eta_\ell = 1$. The quadruple differential distribution can be obtained as 
\begin{equation}
\frac{1}{\sigma} \frac{d\sigma}{d\Omega_1 d\Omega_2^*} = \frac{6}{(4\pi)^2} \sum_{i,j,r,s} (\rho_{tW})_{ir,js} (\Gamma_1)_{ij} (\Gamma_2)_{rs} \,,
\end{equation}
with the indices $i,j=1,2$ corresponding to the top spin space and $r,s=1,2,3$ to the $W$ spin space. The algebra yields
\begin{align}
\frac{1}{\sigma} \frac{d\sigma}{d\Omega_1 d\Omega_2^*} =
\frac{1}{(4\pi)^2} & \left[
1 + a_{m} b_1  Y_1^{m} (\Omega_1) + A_{LM} B_L Y_L^M (\Omega_2^*) \right. \notag \\
& \left. + C_{mLM} b_1 B_L Y_1^m (\Omega_1) Y_L^M (\Omega_2^*) \right] \,,
\label{ec:tWdist}
\end{align}
with $Y_L^M$ the spherical harmonics and
\begin{equation}
b_1 = \alpha \sqrt{\frac{4\pi}{3}} \,,\quad B_1 = - \sqrt{2\pi} \eta_\ell \,,\quad B_2 = \sqrt{\frac{2\pi}{5}} \,.
\end{equation}
For $\bar t W^+$ entanglement the same equations can be used, with $\alpha = -1$ for the negative charged lepton from the $\bar t$ decay and $\eta_\ell = -1$ for the positive lepton from $W^+$ decay.

The density operator for a vector boson pair using the parameterisation of irreducible operators has been written before~\cite{Aguilar-Saavedra:2022wam}. In our case,
\begin{align}
\rho_{WW} =\frac{1}{9}\left[ \mathbb{1}_3\otimes \mathbb{1}_3
+ A^1_{LM}\ T^L_{M} \otimes \mathbb{1}_3 
+ A^2_{LM}\ \mathbb{1}_3\otimes T^L_{M}
+C_{L_1 M_1 L_2 M_2}\ T^{L_1}_{M_1}\otimes T^{L_2}_{M_2}
\right] \,,
\end{align}
where the superindices $1,2$ refer to the $W^+$ and $W^-$ boson, respectively. The corresponding angular distribution is
\begin{align}
\frac{1}{\sigma}\frac{d\sigma}{d\Omega_1^* d\Omega_2^*}
&= \frac{1}{(4\pi)^2}\left[ 1 +A_{LM}^1 B_L^1 Y_L^M(\Omega_1^*) + A_{LM}^2 B_L^2 Y_L^M(\Omega_2^*) \right. \notag \\
& \left. + C_{L_1 M_1 L_2 M_2}  B_{L_1}^1 B_{L_2}^2 Y_{L_1}^{M_1}(\Omega_1^*) Y_{L_2}^{M_2}(\Omega_2^*)  \right] \,,
\label{ec:WWdist}
\end{align}
where in this case the $\ell^+$ three-momentum
\begin{equation}
\hat p_1^* = (\sin \theta_1^* \cos \varphi_1^*, \sin \theta_1^* \sin \varphi_1^*, \cos \theta_1^* )
\end{equation}
is taken in the $W^+$ rest frame.

\section{Calculational setup}
\label{sec:3}

The $t \bar t \to \ell^+ \nu b \ell^- \nu \bar b$ Monte Carlo samples required for our study are generated with {\scshape MadGraph}~\cite{Alwall:2014hca} at the leading order, using NNPDF 3.1~\cite{NNPDF:2017mvq} parton density functions and setting as factorisation and renormalisation scale the average transverse mass, $Q = 1/2 [(m_t^2 + p_{T t}^2)^{1/2} + (m_t^2 + p_{T \bar t}^2)^{1/2} ]$, with $p_T$ the transverse momentum in the usual notation. This is sufficient for our purpose, since next-to-leading order (NLO) corrections to the $t \bar t$ spin correlation coefficients are small~\cite{Bernreuther:2001rq}; the effect of including NLO corrections in entanglement studies has been explicitly tested in Ref.~\cite{Severi:2021cnj}, and found to be small compared to the statistical uncertainty. NLO corrections to the top quark decay have a negligible effect in the charged lepton distributions, changing the value of $\alpha$ at the permille level~\cite{Brandenburg:2002xr}. The total cross section is normalised to the next-to-next-to-leading order prediction~\cite{Czakon:2011xx}.

Two samples are generated, using a centre-of-mass (CM) energy of 13 TeV. A first sample with $t \bar t$ invariant mass $m_{t \bar t} \leq 400$ GeV, containing $2.5 \times 10^7$ events, is used to test $tW$ and $WW$ entanglement near threshold. A second sample with $5 \times 10^6$ events is used to study $tW$ entanglement in the boosted central region. This sample is generated with $m_{t \bar t} \geq 750$ GeV and also with a cut on the scattering angle $\theta_t$ between the top quark momentum in the CM frame and $\hat p_p = (0,0,1)$, $|\cos \theta_t| \leq 0.7$.

We work at the parton level and do not include backgrounds, which are small for the $t \bar t$ dilepton decay channel, especially when the two leptons have different flavour. It is known that for the dilepton decay channel the final state can be reconstructed and the detector effects can be properly accounted for by an unfolding to parton level, as it has already been done by the ATLAS and CMS Collaborations for the measurement of $t \bar t$ spin correlation coefficients~\cite{ATLAS:2016bac,CMS:2019nrx}, using various methods for the reconstruction of the neutrino momenta via kinematic fitting~\cite{D0:1997pjc,CMS:2015rld}. Having this in mind, we use the true top quark and $W$ boson momenta for the computations.
We include an efficiency factor of 0.12 to take into account the detection and reconstruction efficiencies, i.e. that the final state objects are well identified and the reconstructed momenta have good agreement with the expected $t \bar t$ kinematics. This value is the average efficiency found in Ref.~\cite{Severi:2021cnj} with a fast detector simulation, which is smaller than the efficiency of 0.17 obtained in Ref.~\cite{Aguilar-Saavedra:2021ngj}, also with fast simulation. 

The reconstruction and unfolding also introduces a systematic uncertainty in the extracted quantities. The measurement performed by the ATLAS Collaboration~\cite{note} shows a significant modeling uncertainty when converting the particle-level measurement to the parton level near the threshold region, in particular for the $t \bar t$ invariant mass bin used $340 \leq m_{t \bar t} \leq 380$ GeV. As pointed out in Ref.~\cite{Aguilar-Saavedra:2022uye}, the suppression of the $q \bar q$ component with a kinematical cut on the $t \bar t$ velocity in the laboratory frame\footnote{Such variable has already been used by the ATLAS Collaboration in the $t \bar t$ charge asymmetry measurement \cite{ATLAS:2013buu}.} allows to loosen the upper cut on $m_{t \bar t}$ while keeping the $t \bar t$ entanglement, and this might constitute an experimental advantage (at the tree level, raising the upper cut to $m_{t \bar t} \leq 390$ GeV increases the cross section by a factor 1.4). In our sensitivity estimations we include a bulk 10\% systematic uncertainty in our entanglement indicator, namely the lowest eigenvalue of $\rho^{T_2}$, to illustrate the effect of systematic uncertainties arising from reconstruction and unfolding. This figure may be too optimistic for a near-threshold measurement, and in any case a detector-level study is necessary to precisely quantify the systematic uncertainty.

We use two different bases to measure polarisations and spin correlation coefficients. The beamline basis is defined with fixed vectors
\begin{equation}
\hat x = (1,0,0) \,, \quad \hat y = (0,1,0) \,,\quad \hat z = (0,0,1) \,.
\end{equation}
The helicity basis is defined with $\hat z = \hat k$, $\hat x = \hat r$, $\hat y = \hat n$, the K, R and N axes being defined as
\begin{itemize}
\item K-axis (helicity): $\hat k$ is a normalised vector in the direction of the top quark three-momentum in the $t \bar t$ rest frame.
\item R-axis: $\hat r$ is in the production plane and defined as $\hat r = (\hat p_p - \cos \theta_t \hat k)/\sin \theta_t$.
\item N-axis: $\hat n = \hat k \times \hat r$ is orthogonal to the production plane.
\end{itemize}
An alternative definition of the helicity basis can be implemented by introducing sign-flipping factors $\operatorname{sign} \cos \theta_t$ in the definition of the R and N axes~\cite{Bernreuther:2015yna}.
With this sign flip, small values appear for $a_1$ and $A_{11}$, at the percent level, which have little effect in the value of the entanglement indicator $\lambda_1$. 

Using either of these bases, the angles entering Eq.~(\ref{ec:tWdist}) and (\ref{ec:WWdist}) can be defined, and the coefficients can be measured by integration using an appropriate kernel, e.g. for the $tW$ density operator,
\begin{align}
& \int \frac{1}{\sigma} \frac{d\sigma}{d\Omega_1 d\Omega_2^*} Y_1^m(\Omega_1) d\Omega_1 d\Omega_2^* = \frac{b_1}{4\pi} a_m \,, \notag \\
& \int \frac{1}{\sigma} \frac{d\sigma}{d\Omega_1 d\Omega_2^*} Y_L^M(\Omega_2^*) d\Omega_1 d\Omega_2^* = \frac{B_L}{4\pi} A_{LM} \,, \notag \\
& \int \frac{1}{\sigma} \frac{d\sigma}{d\Omega_1 d\Omega_2^*} Y_1^m(\Omega_1) Y_L^M(\Omega_2^*) d\Omega_1 d\Omega_2^* = \frac{b_1 B_L}{(4\pi)^2} C_{mLM} \,.
\end{align}
We remark that these equations are valid even if a kinematical selection is placed on the angle $\theta_W$ between $\vec p_W$ and the $\hat z$ direction, as discussed in the introduction.\footnote{On the other hand, the negative lepton cannot be used as spin analyser for $\bar t$ precisely due to this angular cut.} We can illustrate those arguments numerically, considering $gg \to t \bar t$ with $m_{tt} \leq 370$ GeV and using the beamline basis. In the basis of $\mathcal{H}_A \otimes \mathcal{H}_B$
\begin{equation}
\{ |\oh \, \oh \rangle \,,\; |\oh \, \moh \rangle \,,\; |\moh \, \oh \rangle \,,\;   |\moh \, \moh \rangle \}
\end{equation}
the $t \bar t$ density matrix is, setting to zero entries at the $10^{-3}$ level or below,
\begin{equation}
\rho_{t \bar t} = 
\left( \! \begin{array}{cccc}
0.061 & 0        & 0         & 0 \\
0        & 0.438 & -0.402 & 0 \\
0        & -0.402 & 0.438 & 0 \\
0        & 0         & 0        &  0.062
\end{array} \! \right)
\end{equation}
This matrix has an eigenvector
\begin{equation}
\psi_{t \bar t} = \frac{1}{\sqrt 2} \left[ |\oh \, \moh \rangle -  |\moh \, \oh \rangle \right ]
\end{equation}
with eigenvalue 0.84. That is, the $t \bar t$ pair is nearly produced in a spin-zero singlet, as it is expected close to threshold. Placing a cut $\cos \theta_W \geq 0.9$, the $tW^-$ density matrix is (see Appendix~\ref{sec:a} for notation)
\begin{equation}
\rho_{tW^-} = 
\left( \! \begin{array}{cccccc}
0.055 & 0         & 0        & 0         & 0        &  0 \\
0        & 0.543  & 0        & -0.358 & 0        & 0 \\
0        & 0         & 0.040 & 0         & 0        & 0 \\
0        & -0.358 & 0        &  0.284 & 0        & 0 \\
0        & 0         & 0        & 0         & 0.070 & 0 \\
0        & 0         & 0        & 0         & 0        & 0.008
\end{array} \! \right) \,.
\end{equation}
This matrix has an eigenvector
\begin{equation}
\psi_{tW^-} = 0.818 |\oh \, 0 \rangle - 0.574  |\moh \, 1 \rangle
\end{equation}
with eigenvalue 0.79, in full agreement with Eq.~(\ref{ec:tWvec}).

\section{Sensitivity estimates}
\label{sec:4}

We do not attempt a multi-dimensional optimisation of the sensitivity to $tW^-$ and $W^+ W^-$ entanglement. Instead, we select either of the regions previously used in Ref.~\cite{Aguilar-Saavedra:2022uye} to study the $t \bar t$ entanglement,
\begin{align}
& \text{Threshold:} && m_{t \bar t} \leq 390~\text{GeV} \,,\quad \beta \leq 0.9 \,, \notag \\
& \text{Boosted:}    && m_{t \bar t} \geq 800~\text{GeV} \,,\quad  |\cos \theta_t| \leq 0.6 \,,
\end{align}
with
\begin{equation}
\beta = \left| \frac{p^z_t + p^z_{\bar t}}{E_t + E_{\bar t}} \right|
\end{equation}
being the velocity of the $t \bar t$ pair in the laboratory frame, in obvious notation. We add to these constraints a cut on $\cos \theta_W$, specified below. Each of these regions, defined by a selection on $m_{t \bar t}$, $\beta$ or $\cos \theta_t$, and $\cos \theta_W$, in which the $tW^-$ and $W^+ W^-$ entanglement is measured, will be referred to as `measurement region'.

The selection of the smallest eigenvalue of $\rho^{T_2}$ as entanglement indicator entails a bias because when reconstructing $\rho^{T_2}$ from a finite sample, a negative eigenvalue may arise even for a positive-definite $\rho^{T_2}$, due to statistical fluctuations that cause mismeasurements of the coefficients in the angular distribution. Therefore, the sensitivity to experimentally establish the entanglement is assessed by comparing (i) pseudo-data corresponding to the SM prediction, in a measurement region where there is entanglement; (ii) the separability hypothesis. Pseudo-experiments are performed to obtain numerically the p.d.f. of two quantities:
\begin{itemize}
\item the smallest eigenvalue of $\rho^{T_2}$ in the measurement region, which we label as $\lambda_1^\text{e}$;
\item the smallest eigenvalue $\lambda_1^\text{s}$ of the {\em positive definite} operator $\rho$ in a suitable `calibration region', where the smallest eigenvalue of $\rho$ is nearly zero.
\end{itemize}
The former, when $\lambda_1^\text{e} < 0$, corresponds to the entanglement scenario. The latter is a proxy for the separability hypothesis. The reader may wonder why we do not use for the separability hypothesis the smallest eigenvalue of $\rho^{T_2}$ in a quite different kinematical region such that the state is separable. We believe it is preferable from the experimental point of view to use a calibration region that is as kinematically as close as possible, in $m_{t \bar t}$, $\beta$ and $\cos \theta_t$, to the measurement region. On the other hand, dropping the constraint on $\cos \theta_W$ does not always result in a separable state.

From a large pool of $t \bar t \to \ell^+ \nu b \ell^- \nu \bar b$ events in the measurement region we select a random sample of $N$ events corresponding to the cross section times the assumed luminosity, including the 0.12 efficiency factor previously mentioned. For each sample we calculate the values of the coefficients in the angular distribution, c.f. (\ref{ec:tWdist}) or (\ref{ec:WWdist}), and subsequently we obtain the matrix expression of $\rho^{T_2}$. This matrix is diagonalised numerically and the lowest eigenvalue $\lambda_1^\text{e}$ is obtained. Repeating this procedure $n$ times, we obtain a probability density function (p.d.f) for $\lambda_1^\text{e}$, which is Gaussian to an excellent approximation.

Subsequently, we identify a calibration region with $\vec p_W$ very close to the $\hat z$ axis, in which the density operator $\rho$ has the lowest eigenvalue $\lambda_1^\text{s}$ quite close to zero, typically at the level of few permille. The rest of eigenvalues happen to be similar to those of $\rho^{T_2}$. We perform $n$ pseudo-experiments, selecting a random sample of $N$ events in this calibration region, calculating the values of the coefficients in the angular distribution, obtaining the matrix expression of $\rho$ and calculating its lowest eigenvalue $\lambda_1^\text{s}$. 
With this second set of pseudo-experiments we obtain the p.d.f. of $\lambda_1^\text{s}$ for the {\em positive definite} operator $\rho$, which has a bias towards negative values due to statistical fluctiations. This p.d.f. is very well approximated by a skew-normal distribution. We remark that 
it is essential that the same number $N$ of events per sample is used to calculate the p.d.f.'s of $\lambda_1^\text{e}$ and $\lambda_1^\text{s}$ so that the latter gives an estimation of the bias towards negative values in the former. Clearly, the bias is smaller the larger the statistics, and in some of the examples below it turns out to be unimportant.

\subsection{$tW^-$ entanglement near threshold}
\label{sec:4.1}

For this analysis we select the beamline basis for simplicity, as the helicity basis gives quite the same results. The reason for these bases being equivalent is that the spin-singlet state is rotationally invariant, so the $t \bar t$ spin configuration is the same in either basis. The measurement region is defined as
\begin{equation}
m_{t \bar t} \leq 390~\text{GeV} \,,\quad \beta \leq 0.9 \,,\quad \cos \theta_W \geq 0.3 \,.
\label{ec:cut1}
\end{equation}
The cross section with these cuts is 1.8 pb. We assume a luminosity of 139 fb$^{-1}$, as collected in the Run 2 of the LHC. With this luminosity the number of expected events is $N = 30140$, with the assumed reconstruction efficiency. The calibration region is defined with $\cos \theta_W \geq 0.98$. 

\begin{figure}[htb]
\begin{center}
\begin{tabular}{cc}
\includegraphics[height=5.5cm,clip=]{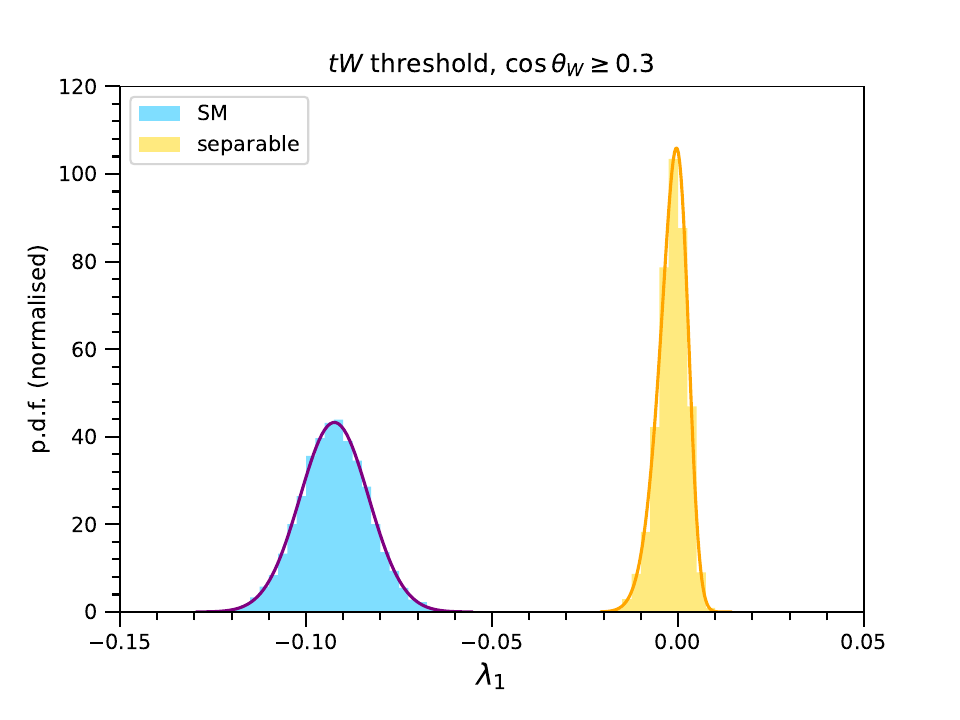}  &
\includegraphics[height=5.5cm,clip=]{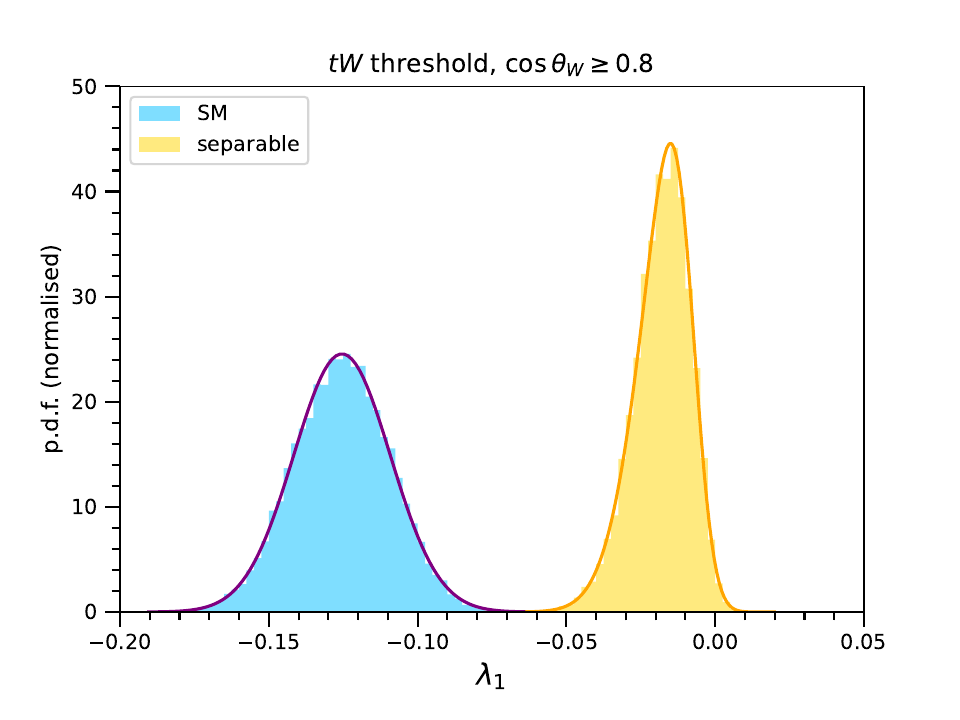} 
\end{tabular}
\caption{Probability density functions of (i) $\lambda_1^\text{e}$ in the threshold measurement region (blue), used to determine $t W^-$ entanglement in the SM; and (ii) $\lambda_1^\text{s}$ in its calibration region (yellow), used as a proxy of the lowest eigenvalue for a separable state.}
\label{fig:thr-B}
\end{center}
\end{figure}

The p.d.f.'s of $\lambda_1^\text{e}$ and $\lambda_1^\text{s}$ obtained for $N = 30140$ events with $n=10^4$ pseudo-experiments are presented on the left panel of Fig.~\ref{fig:thr-B}. In order to better understand the bias issue, we show on the right panel the same for $\cos \theta_W \geq 0.8$, in which case $N = 8870$. The comparison of the two examples shows several illuminating features:
\begin{enumerate}
\item A tighter cut on $\cos \theta_W$ lowers the mean of the distribution $\mu_{\lambda_1^\text{e}}$ --- which we identify with the `measured' value $\lambda_1^\text{e}$ to easy the notation --- but increases the standard deviation $\sigma_{\lambda_1^\text{e}}$, which we identify with the statistical uncertainty on $\lambda_1^\text{e}$. Namely, for $\cos \theta_W \geq 0.3$ we obtain $\lambda_1^\text{e} = -0.092 \pm 0.009$, whereas for  $\cos \theta_W \geq 0.8$ we obtain $\lambda_1^\text{e} = -0.125 \pm  0.016$. 
\item A larger sample shifts the mean of the $\lambda_1^\text{s}$ distribution closer to zero: the bias induced by the finite sample statistics is smaller, as expected.
\end{enumerate}
The optimal cut on $\cos \theta_W$ is a compromise between having a smaller $\lambda_1^\text{e}$, or having a smaller uncertainty and smaller bias. We estimate the significance of a potential measurement with the figure of merit
\begin{equation}
E = \frac{|\lambda_1^\text{e} - \lambda_1^\text{s}|}{\sigma_1} \,, \quad \sigma_1 = \left[ \sigma_{\lambda_1^\text{e}}^2 + (0.1 \lambda_1^\text{e})^2 \right]^\frac{1}{2} \,,
\label{ec:EI}
\end{equation}
where in the estimation of the uncertainty $\sigma_1$ we have added in quadrature the statistical one and a 10\% systematic uncertainty. The numerator $|\lambda_1^\text{e} - \lambda_1^\text{s}|$  corrects for the bias towards negative values. This simple prescription is sufficient because $|\lambda_1^\text{s}|$ is small compared to  $|\lambda_1^\text{e}|$ and  $\sigma_{\lambda_1^\text{e}}$, and also the width of the $\lambda_1^s$ distribution is smaller than $\sigma_{\lambda_1^\text{e}}$.
The denominator of $E$ takes into account a 10\% systematic uncertainty in $\lambda_1^\text{e}$. For the selected region in (\ref{ec:cut1}) we find $E = 7.0$, namely a significance of 7 standard deviations.

\subsection{$tW^-$ entanglement in the boosted region}
\label{sec:4.2}

\begin{figure}[htb]
\begin{center}
\includegraphics[height=6cm,clip=]{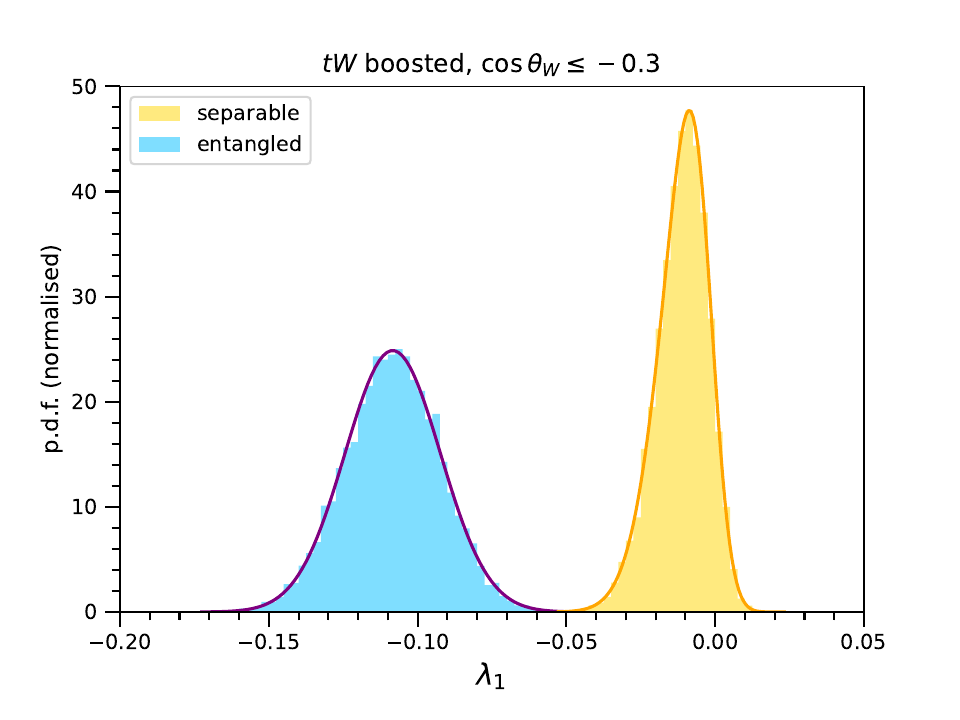}
\caption{Probability density functions of (i) $\lambda_1^\text{e}$ in the boosted measurement region (blue), used to determine $t W^-$ entanglement in the SM; and (ii) $\lambda_1^\text{s}$ in its calibration region (yellow), used as a proxy of the lowest eigenvalue for a separable state.}
\label{fig:boo-H}
\end{center}
\end{figure}

For the boosted region we use the helicity basis. The measurement region is
\begin{equation}
m_{t \bar t} \geq 800~\text{GeV} \,,\quad |\cos \theta_t| \leq 0.6 \,,\quad \cos \theta_W \leq -0.3 \,.
\label{ec:cut2}
\end{equation}
Here we take $\vec p_W$ and the top quark momentum $\hat k$ in opposite hemispheres, so that the $W^-$ boson is more energetic in the laboratory frame. (This may be an advantage from the experimental point of view.) A completely equivalent analysis can be done with $\cos \theta_W \geq 0.3$. 
The cross section with these cuts is 197 fb. We assume a luminosity of 139 fb$^{-1}$ with Run 2 data, and a projection of 250 fb$^{-1}$ at 13.6 TeV in Run 3. With these luminosities the number of expected events is $N = 9800$. The calibration region is $\cos \theta_W \leq -0.9$. The p.d.f.'s of $\lambda_1^\text{e}$ and $\lambda_1^\text{s}$ with $n = 10^4$ pseudo-experiments are shown in Fig.~\ref{fig:boo-H}. The expected sensitivity in the measurement of the lowest eigenvalue of $\rho_{tW}^{T_2}$ is $\lambda_1^\text{e} = -0.108 \pm 0.016$. The figure of merit (\ref{ec:EI}) gives a significance of $5.0\sigma$.

\subsection{$W^+ W^-$ entanglement near threshold}
\label{sec:4.3}

For completeness we also study $W^+ W^-$ entanglement. We point out that the entanglement between $W^+ W^-$ pairs from Higgs boson decays is measurable already with Run 2 data~\cite{Aguilar-Saavedra:2022mpg}. In $t \bar t$ decays the distinct feature is the presence of two additional $b$ quarks, which make necessary the use of special kinematical configurations in which the trace over unmeasured $b$ polarisations does not wash out the entanglement.

Due to the limited statistics we only consider a measurement region near threshold,\begin{equation}
m_{t \bar t} \leq 390~\text{GeV} \,,\quad \beta \leq 0.9 \,,\quad \cos \theta_{W^+} \geq 0.3 \,,\quad \cos \theta_{W^-} \leq -0.3\,,
\label{ec:cut3}
\end{equation}
and use the helicity basis.
The ranges of $\cos \theta_{W^\pm}$ are chosen to have the $W^\pm$ boson momenta roughly aligned with the parent top (anti-)quark momenta in the CM frame, so that the $W^\pm$ momenta in the laboratory frame are larger. (There are three additional configurations that give exactly the same significance.) The cross section with the kinematical selection (\ref{ec:cut3}) is 600 fb. We assume a luminosity of 139 fb$^{-1}$ with Run 2 data, plus 250 fb$^{-1}$ in Run 3. With these luminosities the number of expected events is $N = 29900$. The calibration region is $\cos \theta_{W^+} \geq 0.7$, $\cos \theta_{W^-} \leq -0.7$.

\begin{figure}[htb]
\begin{center}
\includegraphics[height=6cm,clip=]{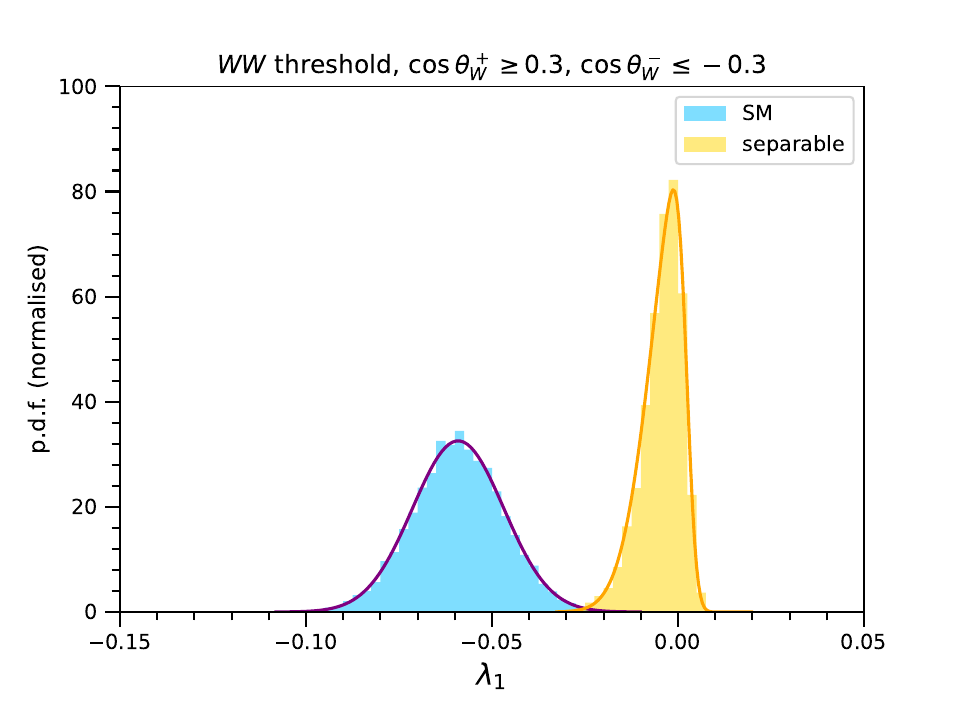}
\caption{Probability density functions of (i) $\lambda_1^\text{e}$ in the boosted measurement region (blue), used to determine $W^+ W^-$ entanglement in the SM; and (ii) $\lambda_1^\text{s}$ in its calibration region (yellow), used as a proxy of the lowest eigenvalue for a separable state.}
\label{fig:WWboo-H}
\end{center}
\end{figure}

The p.d.f.'s of $\lambda_1^\text{e}$ and $\lambda_1^\text{s}$ obtained with $n=10^4$ pseudo-experiments are presented in Fig.~\ref{fig:WWboo-H}. Even if the statistics are comparable to the example in section~\ref{sec:4.1}, the significance is smaller because the central value of the $\lambda_1^\text{e}$ distribution is larger and closer to zero --- which is expected because the coherence is partially lost when considering the $W^+$ boson instead of the top quark. From the pseudo-experiments we find $\lambda_1^\text{e} = -0.0059 \pm 0.004$. The expected significance for the entanglement measurement is $4.0\sigma$.

\section{Discussion}
\label{sec:5}

In this work we have addressed the quantum entanglement involving decay products of $t \bar t$ pairs produced at the LHC, namely between the top quark and the $W^-$ boson from the $\bar t$ decay (or equivalently, between $\bar t$ and $W^+$), and between the two $W$ bosons. 
The key to measure entanglement involving top (anti-)quark decay products is to restrict the angle between the $W$ momentum in the parent top rest frame, thereby avoiding the decoherence caused by the sum over the unmeasured $b$ quark polarisations.

We have investigated, with an analysis at the parton level, the feasibility of several measurements. The estimated sensitivities are collected in Table~\ref{tab:summ}. For these figures, we have included a bulk reconstruction efficiency of 0.12, and assumed a 10\% systematic uncertainty in the entanglement indicator. The sensitivities assuming only statistical uncertainties are also given for reference. We note that several other kinematical regions are possible, which are equivalent from the theoretical point of view, and might be more or less favourable experimentally. For $WW$ entanglement, it is also possible that more sensitive tests exist, since the Peres-Horodecki condition is sufficient but not necessary for 
$\text{dim} \mathcal{H}_A = \text{dim} \mathcal{H}_B = 3$.

\begin{table}[htb]
\begin{center}
\begin{tabular}{cccc}
& & \multicolumn{2}{c}{Expected significance} \\
Measurement            & Luminosity                              & 10\% syst     & stat only \\
$tW^-$ thresold         & 139 fb$^{-1}$                          & $7.0\sigma$ & $9.8\sigma$ \\
$tW^-$ boosted         & 139 fb$^{-1}$ + 250 fb$^{-1}$ & $5.0\sigma$ &  $6.1\sigma$ \\
$W^+W^-$ threshold & 139 fb$^{-1}$ + 250 fb$^{-1}$ & $4.0\sigma$ & $4.5\sigma$
\end{tabular}
\end{center}
\caption{Summary of expected significance for entanglement measurements.}
\label{tab:summ}
\end{table}

A further possibility that could be pursued by experiments to increase the significance is to combine disjoint regions into an entanglement measurement. For example, for $tW^-$ entanglement one could perform two measurements, with $\cos \theta_W \geq 0.3$, and $\cos \theta_W \leq -0.3$, and combine them in order to gain statistics. This type of combination with a proper accounting of systematic uncertainties and correlations, can only be performed by an experiment.

In conclusion, $t \bar t$ production offers a rare possibility of measuring spin entanglement between a boson and a fermion, which would also constitute the first measurement at the energy frontier. And this would be possible with the data already collected at the LHC Run 2.

\section*{Acknowledgements}

I thank A. Casas and J. Moreno for useful discussions. This work of has been supported by MICINN projects PID2019-110058GB-C21, PID2022-142545NB-C21 and CEX2020-001007-S funded by MCIN/AEI/10.13039/501100011033 and by ERDF, and by Funda\c{c}{\~a}o para a Ci{\^e}ncia e a Tecnologia (FCT, Portugal) through the project CERN/FIS-PAR/0019/2021.

\appendix
\section{Explicit expressions for $tW$ density operators}
\label{sec:a}

In the basis of the product space $\mathcal{H}_A \otimes \mathcal{H}_B$
\begin{equation}
\{ |\oh \, 1 \rangle \,,\; |\oh \, 0 \rangle \,,\; |\oh \, \shm\! 1 \rangle \,,\;  
 |\moh \, 1 \rangle \,,\; |\moh \, 0 \rangle \,,\; |\moh \, \shm \! 1 \rangle \}  
\end{equation}
the matrix elements of the density operator $\rho_{tW}$ are
\begin{align}
& (\rho_{tW})_{11} = \frac{1}{6} \left[ 1+a_0 \right]  
+ \frac{1}{2 \sqrt 6} \left[ A_{10} +  C_{010} \right]
+ \frac{1}{6 \sqrt 2} \left[  A_{20} + C_{020} \right] 
\,, \notag \\
& (\rho_{tW})_{12} = -\frac{1}{2 \sqrt{6}} \left[ A_{11} + A_{21} + C_{011} + C_{021}  \right]
\,, \notag \\
& (\rho_{tW})_{13} = \frac{1}{2 \sqrt{3}} \left[ A_{22} + C_{022}  \right]
\,, \notag \\
& (\rho_{tW})_{14} = - \frac{1}{3 \sqrt 2} a_1 - \frac{1}{2 \sqrt{3}}  C_{110} - \frac{1}{6} C_{120}
\,, \notag \\
& (\rho_{tW})_{15} = \frac{1}{2 \sqrt{3}} \left[  C_{111} + C_{121} \right]
\,, \notag \\  
& (\rho_{tW})_{16} = -\frac{1}{\sqrt{6}}  C_{122}
\,, \notag \\
& (\rho_{tW})_{22} = \frac{1}{6}  \left[ 1+a_0 \right]  - \frac{1}{3 \sqrt 2} \left[ A_{20} + C_{020} \right]
\,, \notag \\
& (\rho_{tW})_{23} = \frac{1}{2 \sqrt{6}} \left[  - A_{11} + A_{21} - C_{011} + C_{021}  \right]
\,, \notag \\  \displaybreak 
& (\rho_{tW})_{24} = -\frac{1}{2 \sqrt{3}} \left[ C_{11\,\shortminus 1} + C_{12\,\shortminus 1} \right]
\,, \notag \\ 
& (\rho_{tW})_{25} = - \frac{1}{3 \sqrt 2} a_1 + \frac{1}{3}  C_{120}
\,, \notag \\ 
& (\rho_{tW})_{26} = \frac{1}{2 \sqrt{3}} \left[  C_{111} - C_{121} \right]
\,, \notag \\
& (\rho_{tW})_{33} = \frac{1}{6} \left[ 1+a_0 \right]  
- \frac{1}{2 \sqrt 6} \left[ A_{10} +  C_{010} \right]
 + \frac{1}{6 \sqrt 2} \left[  A_{20} + C_{020} \right]
\,, \notag \\ 
& (\rho_{tW})_{34} = -\frac{1}{\sqrt{6}}  C_{12\,\shortminus 2} 
\,, \notag \\ 
& (\rho_{tW})_{35} = \frac{1}{2 \sqrt{3}}  \left[ C_{12\,\shortminus 1} - C_{11\,\shortminus 1} \right]
\,, \notag \\
& (\rho_{tW})_{36} = - \frac{1}{3 \sqrt 2} a_1 + \frac{1}{2 \sqrt{3}}  C_{110} - \frac{1}{6} C_{120}
\,, \notag \\
& (\rho_{tW})_{44} = \frac{1}{6} \left[ 1-a_0 \right]  
 + \frac{1}{2 \sqrt 6} \left[ A_{10} -  C_{010} \right]
  + \frac{1}{6 \sqrt 2} \left[  A_{20} - C_{020} \right]
\,, \notag \\
& (\rho_{tW})_{45} = \frac{1}{2 \sqrt{6}}  \left[ - A_{11} - A_{21} + C_{011} + C_{021} \right] 
\,, \notag \\
& (\rho_{tW})_{46} = \frac{1}{2 \sqrt{3}}  \left[ A_{22} - C_{022} \right]
\,, \notag \\
& (\rho_{tW})_{55} = \frac{1}{6} \left[ 1-a_0 \right]  
- \frac{1}{3 \sqrt 2} \left[ A_{20} - C_{020} \right]
\,, \notag \\
& (\rho_{tW})_{56} = \frac{1}{2 \sqrt{6}}  \left[ - A_{11} + A_{21} + C_{011} - C_{021} \right]
\,, \notag \\
& (\rho_{tW})_{66} =  \frac{1}{6} \left[ 1-a_0 \right]  
- \frac{1}{2 \sqrt 6} \left[ A_{10} -  C_{010} \right]
  + \frac{1}{6 \sqrt 2} \left[  A_{20} - C_{020} \right] \,.
\end{align}
The operator $\rho_{tW}^{T_2}$ has matrix elements $(\rho_{tW}^{T_2})_{ii} =  (\rho_{tW})_{ii}$, $(\rho_{tW}^{T_2})_{j \; j+3} = (\rho_{tW})_{j \; j+3}$, for $i=1,\dots,6$, $j=1,2,3$, and
\begin{align}
& (\rho_{tW}^{T_2})_{12} = \frac{1}{2 \sqrt{6}}  \left[ A_{1\,\shm 1} + A_{2\,\shm 1} + C_{01\,\shm 1} + C_{02\,\shm 1}  \right]
\,, \notag \\
& (\rho_{tW}^{T_2})_{13} = \frac{1}{2 \sqrt{3}}  \left[  A_{2\,\shm 2} + C_{02\,\shm 2} \right]
\,, \notag \\
& (\rho_{tW}^{T_2})_{15} = -\frac{1}{2 \sqrt{3}}  \left[  C_{11\,\shm 1} + C_{12\,\shm 1} \right]
\,, \notag \\
& (\rho_{tW}^{T_2})_{16} = -\frac{1}{\sqrt{6}}  C_{12\,\shm 2}
\,, \notag \\
& (\rho_{tW}^{T_2})_{23} = \frac{1}{2 \sqrt{6}}  \left[ A_{1\,\shm 1} - A_{2\,\shm 1}  + C_{01\,\shm 1} - C_{02\,\shm 1} \right]
\,, \notag \\ 
& (\rho_{tW}^{T_2})_{24} = \frac{1}{2 \sqrt{3}}  \left[ C_{111} + C_{121} \right]
\,, \notag \\ 
& (\rho_{tW}^{T_2})_{26} = \frac{1}{2 \sqrt{3}}  \left[  C_{12\,\shm 1} - C_{11\,\shm 1} \right]
\,, \notag \\ 
& (\rho_{tW}^{T_2})_{34} = -\frac{1}{\sqrt{6}}  C_{122} 
\,, \notag \\ \displaybreak 
& (\rho_{tW}^{T_2})_{35} = \frac{1}{2 \sqrt{3}}  \left[ C_{111} - C_{121} \right]
\,, \notag \\ & (\rho_{tW}^{T_2})_{45} = \frac{1}{2 \sqrt{6}}  \left[ + A_{1\,\shm 1} + A_{2\,\shm 1} - C_{01\,\shm 1} - C_{02\,\shm 1}  \right]
\,, \notag \\
& (\rho_{tW}^{T_2})_{46} = \frac{1}{2 \sqrt{3}}  \left[ A_{2\,\shm 2} - C_{02 \shm -2} \right]
\,, \notag \\
& (\rho_{tW}^{T_2})_{56} = \frac{1}{2 \sqrt{6}}  \left[ A_{1\,\shm 1} - A_{2\,\shm 1}  -C_{01\,\shm 1} + C_{02\,\shm 1} \right] \,.
\end{align}


\begin{thebibliography}{99}


\bibitem{Afik:2020onf}
Y.~Afik and J.~R.~M.~de Nova,
``Entanglement and quantum tomography with top quarks at the LHC,''
Eur. Phys. J. Plus \textbf{136}, no.9, 907 (2021)
[arXiv:2003.02280 [quant-ph]].

\bibitem{Fabbrichesi:2021npl}
M.~Fabbrichesi, R.~Floreanini and G.~Panizzo,
``Testing Bell Inequalities at the LHC with Top-Quark Pairs,''
Phys. Rev. Lett. \textbf{127} (2021) no.16, 16
[arXiv:2102.11883 [hep-ph]].

\bibitem{Severi:2021cnj}
C.~Severi, C.~D.~Boschi, F.~Maltoni and M.~Sioli,
``Quantum tops at the LHC: from entanglement to Bell inequalities,''
Eur. Phys. J. C \textbf{82}, no.4, 285 (2022)
[arXiv:2110.10112 [hep-ph]].

\bibitem{Afik:2022kwm}
Y.~Afik and J.~R.~M.~de Nova,
``Quantum information with top quarks in QCD,''
Quantum \textbf{6}, 820 (2022)
[arXiv:2203.05582 [quant-ph]].

\bibitem{Aguilar-Saavedra:2022uye}
J.~A.~Aguilar-Saavedra and J.~A.~Casas,
``Improved tests of entanglement and Bell inequalities with LHC tops,''
Eur. Phys. J. C \textbf{82}, no.8, 666 (2022)
[arXiv:2205.00542 [hep-ph]].

\bibitem{Afik:2022dgh}
Y.~Afik and J.~R.~M.~de Nova,
``Quantum Discord and Steering in Top Quarks at the LHC,''
Phys. Rev. Lett. \textbf{130}, no.22, 221801 (2023)
[arXiv:2209.03969 [quant-ph]].

\bibitem{Dong:2023xiw}
Z.~Dong, D.~Gon\c{c}alves, K.~Kong and A.~Navarro,
``When the Machine Chimes the Bell: Entanglement and Bell Inequalities with Boosted $t\bar{t}$,''
[arXiv:2305.07075 [hep-ph]].

\bibitem{note}
ATLAS Collaboration, 
``Observation of quantum entanglement in top-quark pair production using $pp$ collisions of $\sqrt s = 13$ TeV with the ATLAS detector,''
report ATLAS-CONF-2023-069.

\bibitem{Barr:2021zcp}
A.~J.~Barr,
``Testing Bell inequalities in Higgs boson decays,''
Phys. Lett. B \textbf{825}, 136866 (2022)
[arXiv:2106.01377 [hep-ph]].

\bibitem{Aguilar-Saavedra:2022wam}
J.~A.~Aguilar-Saavedra, A.~Bernal, J.~A.~Casas and J.~M.~Moreno,
``Testing entanglement and Bell inequalities in $H \to \to ZZ$,''
Phys. Rev. D \textbf{107}, no.1, 016012 (2023)
[arXiv:2209.13441 [hep-ph]].

\bibitem{Aguilar-Saavedra:2022mpg}
J.~A.~Aguilar-Saavedra,
``Laboratory-frame tests of quantum entanglement in $H \to WW$,''
Phys. Rev. D \textbf{107}, no.7, 076016 (2023)
[arXiv:2209.14033 [hep-ph]].



\bibitem{Ashby-Pickering:2022umy}
R.~Ashby-Pickering, A.~J.~Barr and A.~Wierzchucka,
``Quantum state tomography, entanglement detection and Bell violation prospects in weak decays of massive particles,''
JHEP \textbf{05}, 020 (2023)
[arXiv:2209.13990 [quant-ph]].

\bibitem{Fabbrichesi:2023cev}
M.~Fabbrichesi, R.~Floreanini, E.~Gabrielli and L.~Marzola,
``Bell inequalities and quantum entanglement in weak gauge bosons production at the LHC and future colliders,''
[arXiv:2302.00683 [hep-ph]].

\bibitem{Morales:2023gow}
R.~A.~Morales,
``Exploring Bell inequalities and quantum entanglement in vector boson scattering,''
[arXiv:2306.17247 [hep-ph]].


\bibitem{Altakach:2022ywa}
M.~M.~Altakach, P.~Lamba, F.~Maltoni, K.~Mawatari and K.~Sakurai,
``Quantum information and CP measurement in $H \to \tau^+ \tau^-$ at future lepton colliders,''
Phys. Rev. D \textbf{107}, no.9, 093002 (2023)
[arXiv:2211.10513 [hep-ph]].


\bibitem{Feist:2022zwe}
A.~Feist, G.~Huang, G.~Arend, Y.~Yang, J.~W.~Henke, A.~S.~Raja, F.~J.~Kappert, R.~N.~Wang, H.~Louren\c{c}o-Martins and Z.~Qiu, \textit{et al.}
``Cavity-mediated electron-photon pairs,''
Science \textbf{377}, no.6607, abo5037 (2022)
[arXiv:2202.12821 [quant-ph]].

\bibitem{Peres:1996dw}
A.~Peres,
``Separability criterion for density matrices,''
Phys. Rev. Lett. \textbf{77}, 1413-1415 (1996)
[arXiv:quant-ph/9604005 [quant-ph]].

\bibitem{Horodecki:1997vt}
P.~Horodecki,
``Separability criterion and inseparable mixed states with positive partial transposition,''
Phys. Lett. A \textbf{232}, 333 (1997)
[arXiv:quant-ph/9703004 [quant-ph]].

\bibitem{Aguilar-Saavedra:2022kgy}
J.~A.~Aguilar-Saavedra,
``Crafting polarizations for top, $W$, and $Z$,''
Phys. Rev. D \textbf{106}, no.11, 115021 (2022)
[arXiv:2208.00424 [hep-ph]].

\bibitem{Aguilar-Saavedra:2015yza}
J.~A.~Aguilar-Saavedra and J.~Bernabeu,
``Breaking down the entire W boson spin observables from its decay,''
Phys. Rev. D \textbf{93}, no.1, 011301 (2016)
[arXiv:1508.04592 [hep-ph]].

\bibitem{Rahaman:2021fcz}
R.~Rahaman and R.~K.~Singh,
``Breaking down the entire spectrum of spin correlations of a pair of particles involving fermions and gauge bosons,''
Nucl. Phys. B \textbf{984}, 115984 (2022)
[arXiv:2109.09345 [hep-ph]].


\bibitem{Alwall:2014hca}
J.~Alwall, R.~Frederix, S.~Frixione, V.~Hirschi, F.~Maltoni, O.~Mattelaer, H.~S.~Shao, T.~Stelzer, P.~Torrielli and M.~Zaro,
``The automated computation of tree-level and next-to-leading order differential cross sections, and their matching to parton shower simulations,''
JHEP \textbf{07}, 079 (2014)
[arXiv:1405.0301 [hep-ph]].

\bibitem{NNPDF:2017mvq}
R.~D.~Ball \textit{et al.} [NNPDF],
``Parton distributions from high-precision collider data,''
Eur. Phys. J. C \textbf{77}, no.10, 663 (2017)
[arXiv:1706.00428 [hep-ph]].

\bibitem{Bernreuther:2001rq}
W.~Bernreuther, A.~Brandenburg, Z.~G.~Si and P.~Uwer,
``Top quark spin correlations at hadron colliders: Predictions at next-to-leading order QCD,''
Phys. Rev. Lett. \textbf{87}, 242002 (2001)
[arXiv:hep-ph/0107086 [hep-ph]].

\bibitem{Brandenburg:2002xr}
A.~Brandenburg, Z.~G.~Si and P.~Uwer,
``QCD corrected spin analyzing power of jets in decays of polarized top quarks,''
Phys. Lett. B \textbf{539}, 235-241 (2002)
[arXiv:hep-ph/0205023 [hep-ph]].

\bibitem{Czakon:2011xx}
M.~Czakon and A.~Mitov,
``Top++: A Program for the Calculation of the Top-Pair Cross-Section at Hadron Colliders,''
Comput. Phys. Commun. \textbf{185}, 2930 (2014)
[arXiv:1112.5675 [hep-ph]].

\bibitem{ATLAS:2016bac}
M.~Aaboud \textit{et al.} [ATLAS Collaboration],
``Measurements of top quark spin observables in $ t\overline{t} $ events using dilepton final states in $ \sqrt{s}=8 $ TeV pp collisions with the ATLAS detector,''
JHEP \textbf{03}, 113 (2017)
[arXiv:1612.07004 [hep-ex]].

\bibitem{CMS:2019nrx}
A.~M.~Sirunyan \textit{et al.} [CMS Collaboration],
``Measurement of the top quark polarization and $\mathrm{t\bar{t}}$ spin correlations using dilepton final states in proton-proton collisions at $\sqrt{s} =$ 13 TeV,''
Phys. Rev. D \textbf{100}, no.7, 072002 (2019)
[arXiv:1907.03729 [hep-ex]].

\bibitem{D0:1997pjc}
B.~Abbott \textit{et al.} [D0],
``Measurement of the top quark mass using dilepton events,''
Phys. Rev. Lett. \textbf{80}, 2063-2068 (1998)
[arXiv:hep-ex/9706014 [hep-ex]].

\bibitem{CMS:2015rld}
V.~Khachatryan \textit{et al.} [CMS],
``Measurement of the differential cross section for top quark pair production in pp collisions at $\sqrt{s} = 8\,\text {TeV} $,''
Eur. Phys. J. C \textbf{75}, no.11, 542 (2015)
[arXiv:1505.04480 [hep-ex]].



\bibitem{Aguilar-Saavedra:2021ngj}
J.~A.~Aguilar-Saavedra, M.~C.~N.~Fiolhais, P.~Mart\'\i{}n-Ramiro, J.~M.~Moreno and A.~Onofre,
``A template method to measure the $t {\bar{t}}$ polarisation,''
Eur. Phys. J. C \textbf{82}, no.2, 134 (2022)
[arXiv:2111.10394 [hep-ph]].

\bibitem{ATLAS:2013buu}
G.~Aad \textit{et al.} [ATLAS Collaboration],
``Measurement of the top quark pair production charge asymmetry in proton-proton collisions at $\sqrt{s}$ = 7 TeV using the ATLAS detector,''
JHEP \textbf{02}, 107 (2014)
[arXiv:1311.6724 [hep-ex]].

\bibitem{Bernreuther:2015yna}
W.~Bernreuther, D.~Heisler and Z.~G.~Si,
``A set of top quark spin correlation and polarization observables for the LHC: Standard Model predictions and new physics contributions,''
JHEP \textbf{12}, 026 (2015)
[arXiv:1508.05271 [hep-ph]].

\end{thebibliography}
\end{document}